\documentclass[
superscriptaddress,
twocolumn,
 amsmath,amssymb,
 aps, physrev,
]{revtex4-2}
\usepackage{orcidlink}
\usepackage[USenglish,english,american]{babel}
\usepackage{graphicx}
\usepackage{xcolor}
\usepackage{dcolumn}
\usepackage{bm}
\usepackage{hyperref}
\usepackage{float}

\begin{document}

\preprint{APS/123-QED}

\title{Tuning ergodicity breaking: Anomalous diffusion under asymptotic power-law forcing}

\author{Raul V. B. Mor\'{a}s}
\affiliation{Instituto Latino-Americano de Ci\^{e}ncias da Vida e da Natureza, Universidade Federal da Integra\c{c}\~ao Latino-Americana, 85867-970 Foz do Igua\c{c}u, Paran\'{a}, Brazil.}

\author{M. Florencia Carusela\orcidlink{0000-0002-7200-8752}}
\affiliation{Instituto de Ciencias, Universidad Nacional de General Sarmiento, Los Polvorines, Buenos Aires, Argentina.
}
\affiliation{National Scientific and Technical Research Council, Argentina.}
\author{Luciano C. Lapas\orcidlink{0000-0002-0029-5625}}
\email{Contact author: luciano.lapas@unila.edu.br}
\affiliation{Instituto Latino-Americano de Ci\^{e}ncias da Vida e da Natureza, Universidade Federal da Integra\c{c}\~ao Latino-Americana, 85867-970 Foz do Igua\c{c}u, Paran\'{a}, Brazil.}

\date{\today}

\begin{abstract}
In non-Markovian systems, distinct dynamical phases arise from the competition between internal memory and external forcing, encompassing thermalization, persistent ergodicity breaking, and runaway energy growth. This study shows that the scaling parameter $\eta$ governs the emergent phase diagram within a system described by the Generalized Langevin Equation, particularly when subjected to external drives with asymptotic power-law tails. Three universal regimes for diffusive processes are delineated by this parameter: thermalization ($\eta > 0$), non-ergodic saturation ($\eta = 0$), and a force-dominated runaway phase ($\eta < 0$). The fluctuation-dissipation theorem, within this framework, is shown to be independent of external force and determined by the integral of noise density of states. A selective breaking of ergodicity is revealed by this formulation; microscopic fluctuations are decoupled from the drive, yet the relaxation completely encodes it, which in turn controls the kinetic effective temperature. Direct Langevin simulations in the Markovian limit quantitatively confirm this classification, capturing the non-thermal plateau at the critical point.
\end{abstract}

\maketitle

{\it Introduction. ---}\label{sec:introduction} Non-equilibrium anomalous diffusion underpins transport in soft matter, biological systems, and driven quantum assemblies~\cite{Metzler2000,Bouchaud1990,Hofling2013}. A fundamental open question is whether an external drive can deterministically tune such systems across the sharp boundary separating ergodic thermalization from persistent ergodicity breaking. Because the stationary state is set by the competition between internal memory and external spectral content at low frequencies, the decisive quantity is the algebraic exponent of the drive’s long-time tail, rather than its functional form. We focus on drives with asymptotic power-law tails, corresponding via Tauberian theorems to the Laplace scaling $\tilde{F}_{\text{ext}}(z) \sim z^{\beta}$ as $z \to 0$. This asymptotic condition encompasses diverse scenarios, from anomalous charge transport in disordered semiconductors~\cite{Metzler2000} to viscoelastic stress relaxation~\cite{Rigato2017} and active cytoskeletal fluctuations~\cite{Goychuk2009,Gallet2009}.

The undriven Generalized Langevin Equation (GLE) already exhibits rich phenomenology: ergodicity breaking is governed by the low-frequency scaling of the memory kernel~\cite{Morgado2002}, a mechanism that dictates the validity of both Khinchin’s Theorem (KT) and the Fluctuation-Dissipation Theorem (FDT)~\cite{Costa2003,Lapas2007,Lapas2008}. This framework has recently been extended to aging dynamics~\cite{Agrawal2019,Gomes-Filho2025} and non-Markovian open quantum systems~\cite{Manikandan2019,Chetrite2021}. Nonetheless, it is presently unclear how a tunable power-law drive impacts this entire phase diagram, nor whether ergodicity and the FDT can be collectively regulated by it.

Here, we derive exact, closed-form asymptotic expressions for the long-time relaxation function and effective temperature of the driven GLE. We show that the interplay between memory-induced dissipation, quantified by the diffusion exponent $\alpha$, and external forcing, quantified by $\beta$, is fully encapsulated by a singular scaling parameter: $\eta=\beta-\alpha+2$. This parameter unambiguously delineates three distinct dynamical phases: thermalization ($\eta > 0$), critical ergodicity breaking ($\eta = 0$), and runaway energy growth ($\eta < 0$). We further show that the FDT extends to a generalized form whose effective bath coupling is fixed entirely by the spectral integral of the intrinsic noise density of states so that microscopic fluctuations remain strictly decoupled from the drive, while the relaxation function and effective temperature are governed by the asymptotic $\alpha$-$\beta$ interplay. Finally, in the Markovian limit ($\eta=\beta+1$), we validate this universal phase criterion against exact Langevin simulations.

{\it Generalized Langevin formalism. ---}\label{sec:GLFormalism} Anomalous diffusion arises when multiple relaxation times compete, rendering initial conditions decisive for the long-time dynamics~\cite{Gomes-Filho2025,Vilar2001}. Such intrinsically non-Markovian processes are described by the GLE,
\begin{equation}
    \frac{dA}{dt} = -\int_0^t \Pi(t-t’) A(t’)\,dt’ + \xi(t) + F_{\text{ext}}(t),
    \label{eq:GLE}
\end{equation}
where $A(t)$ is a dynamical variable (e.g., momentum), $\Pi(t)$ the memory kernel, and $F_{\text{ext}}$ an external drive. For $\Pi(t)=2\gamma\delta(t)$, Eq.~\eqref{eq:GLE} recovers the ordinary Langevin equation. The stochastic force $\xi(t)$ is defined by a Gaussian distribution, exhibiting a zero ensemble average, $\langle\xi(t)\rangle=0$, and the FDT is expressible as
\begin{equation}
    C_\xi(t-t’)=\langle\xi(t)\xi(t’)\rangle=\phi(t-t’),
    \label{eq:C_xi-1}
\end{equation}
with $\phi(t-t')$ the time-lag autocorrelation function; causality requires $\langle A(0)\xi(t)\rangle=0$. The Laplace transform of the GLE gives $\tilde{A}(z) = \big[A(0)+\tilde{\xi}(z)+\tilde{F}_{\text{ext}}(z)\big]\tilde{R}(z)$, with relaxation function
\begin{equation}
    \tilde{R}(z)=\frac{1}{z+\tilde{\Pi}(z)}.
    \label{TL:R}
\end{equation}

{\it Anomalous diffusion without external force ---}\label{subsec:ADwithoutExtForce} Defining the force-subtracted variable $A^{\dag}(t)=A(t)-\int_0^t R(t-t’)F_{\text{ext}}(t’)\,dt’$ yields the homogeneous dynamics
\begin{equation}
    \frac{dA^{\dag}}{dt}=-\int_0^t \Pi(t-t’)A^{\dag}(t’)\,dt’+\xi(t).
    \label{eq:GLE2}
\end{equation}
For the unforced GLE, the response function satisfies $\dot{R}(t)=-\int_0^t \Pi(t-t’) R(t’) dt’$. We can write the relaxation function over time as $R(t)=C_{A^{\dag}}(t)/C_{A^{\dag}}(0)$, where $C_{A^{\dag}}(t)$ is the autocorrelation function of $A^{\dag}(t)$. Employing the Final Value Theorem (FVT), the long-time behavior can be deduced from the small-$z$ limit. Following~\cite{Morgado2002}, we write the kernel in the asymptotic form 
\begin{equation}
\tilde{\Pi}(z)\sim c_T z^{\alpha-1},
\label{eq:morgado}
\end{equation}
where $c_T$ is a positive constant that absorbs the prefactor dependent on the microscopic details of the system. By inserting this into Eq.~\eqref{TL:R} yields
\begin{equation}
R_\infty\equiv \lim_{t\to\infty}R(t)=\lim_{z\to 0}z\tilde{R}(z)=\lim_{z \to 0}\frac{1}{ 1+c_T z^{\alpha-2}}.
\label{eq:R(t->infy)-TVF}
\end{equation}
In addition, the kinetic effective temperature reads $T_{\text{eff}}(t)=T_0+(T_0-T_R)\,[R^2(t)-1]$, with $T_0$ and $T_R$ the initial and reservoir temperatures~\cite{Lapas2007}. The system’s thermalization with the thermal reservoir is guaranteed, irrespective of initial conditions, due to the fulfillment of the irreversibility (or mixing) condition ($R_\infty=0$). This condition is characteristic of all diffusive processes where $0<\alpha < 2$. Nevertheless, if $C_{A^\dag}(t)$ persists non-zero over an extended period, the system will not undergo mixing and consequently will not thermalize, resulting in $\lim_{t\to \infty}T_{\text{eff}}(t)\neq T_R$. This is the case of ballistic diffusion, $\alpha = 2$, where $R_\infty = (1 + c_T)^{-1}$; the ballistic system settles into a state and persists in it, exhibiting a residual current $\langle A^\dag\rangle$ (a \textit{drift}). Despite the failure of the irreversibility condition and the consequent violation of the Ergodic Hypothesis (EH), KT remains valid~\cite{Lapas2008}.

In Fourier space, Eq.~\eqref{eq:C_xi-1} becomes
$\langle \xi(\omega) \xi(\omega’) \rangle = D_T \Pi(\omega - \omega^{\prime})$ under $\phi(t)=D_T\Pi(t)$, with $\Pi(\omega-\omega’)$ the frequency-domain memory function. For a stationary ensemble~\cite{Kubo1966}, $\langle \xi_{\omega}\xi_{\omega^{\prime}}\rangle = 2\pi S(\omega) \delta(\omega + \omega^{\prime})$, and inverting using the parity of $S(\omega)$ gives
\begin{equation}
  \Pi(t) = \int_0^{\infty} d\omega \rho(\omega) \cos(\omega t),
  \label{imp-gamma-rho}
\end{equation}
where the Noise Density of States (NDS) $\rho (\omega) \equiv S(\omega)/(\pi D_T)$ is an intrinsic reservoir property; in the Caldeira-Leggett framework~\cite{Caldeira1983}, it arises from the bath spectral density, since tracing over the oscillator bath yields a GLE whose memory kernel is fixed by $\rho(\omega)$ alone. Then $\tilde \Pi(z) = \int_0^\infty d\omega  z \rho(\omega)/(z^2+\omega^2)$, and the FVT applied to Eq.~\eqref{TL:R} fixes the long-time relaxation,
\begin{equation}
  R_\infty = \frac{1}{1+M_\omega},
  \qquad
  M_\omega = \int_0^\infty d\omega\,\frac{\rho(\omega)}{\omega^2}.
  \label{eq:limR-2}
\end{equation}
Because $M_\omega$ encodes the bath structure alone, this central result ties the relaxation function (and hence the effective temperature) entirely to intrinsic reservoir properties, establishing a microscopic link between fluctuations, dissipation, and thermalization.

{\it \label{subsec:ADwithExtForce} Anomalous diffusion with external force ---} The relaxation function, when an external drive is involved, is described as
\begin{equation}
    R_{\text{ext}}(t) = \frac{C_A(t)}{C_A(0)}
    = R(t) + \epsilon \int_0^t R(t-t’)\,F_{\text{ext}}(t’)\,dt’,
    \label{eq:Rext}
\end{equation}
where $\epsilon=\langle A(0)\rangle/\langle A^2(0)\rangle$ quantifies initial conditions. For $F_{\text{ext}}=0$ or $\langle A(0)\rangle=0$, $R_{\text{ext}}(t)=R(t)$. Otherwise, the Laplace transform together with Eq.~\eqref{TL:R} gives

\begin{equation}
    \tilde{R}_{\text{ext}}(z) = \frac{1+\epsilon\tilde{F}_{\text{ext}}(z)}{  z + \tilde{\Pi}(z)}.
    \label{eq:R(z)_final}
\end{equation}

\begin{figure}[t]
    \centering
    \includegraphics[width=0.95\linewidth]{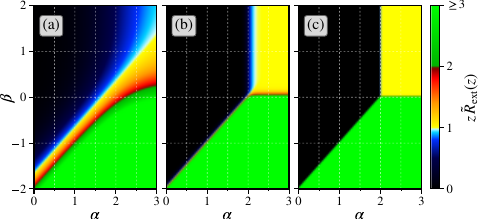}
    \caption{
        Asymptotic relaxation function $z\tilde{R}_{\text{ext}}(z)$ in the $\alpha$--$\beta$ plane. \textbf{(a)} $z=10^{-1}$ and \textbf{(b)} $z=10^{-10}$ are transient regimes; \textbf{(c)} $z=10^{-100}$ is the long-time regime, where $z$ denotes the dimensionless Laplace variable.
    }
    \label{fig:beta_alfa_z}
\end{figure}

To examine the long-time behavior, we consider a general drive $F_{\text{ext}}(t)$ whose Laplace transform admits the small-$z$ asymptotic expansion
\begin{equation}
\tilde{F}_{\text{ext}}(z) \sim k z^{\beta},
\label{eq:asympF}
\end{equation}
with $k$ a characteristic amplitude. The correspondence between the
small-$z$ limit in Laplace space and the long-time asymptotic regime is
not merely a formal inversion, but a rigorous consequence of Tauberian
theorems~\cite{Korevaar2004}: if $\tilde{F}_{\text{ext}}(z)\sim z^{\beta}$
as $z\to 0$ under suitable regularity, then
$F_{\text{ext}}(t) \sim t^{-\beta-1}/\Gamma(-\beta)$, so a power-law tail
in time reflects a branch-point singularity at the origin. The exponent
$\beta$ thus characterizes the drive entirely through its asymptotic
tail, independently of its detailed waveform. Substituting the asymptotic
forms into Eq.~\eqref{eq:R(z)_final} and applying the FVT yields
\begin{equation}
R_{\text{ext},\infty} \sim \frac{1 + \epsilon k z^{\beta}}{1 + c_T z^{\alpha - 2}} ,
\label{eq:limR(t)}
\end{equation}
expressing the long-time response in terms of the small-$z$ asymptotics of
the memory kernel and the drive.

Figure~\ref{fig:beta_alfa_z} maps $\lim_{z\to 0} z\tilde{R}_{\text{ext}}(z)$ over the $(\alpha,\beta)$-plane for fixed $z$ [(a) $10^{-1}$, (b) $10^{-10}$, (c) $10^{-100}$], with $\epsilon k = 2$ and $c_T = 1$. The transient panel (a) is less accurate, as subleading orders of $\tilde\Pi$ and $\tilde F_{\text{ext}}$ are neglected; precision improves as $z\to 0$, and panel (c) shows sharp contours consistent with a transition between diffusive and hydrodynamic processes~\cite{Vainstein2006,Agrawal2019}. In the color scale, black marks the regime in which the irreversibility condition is satisfied ($R_{\text{ext},\infty}=0$), thus the KT, the FDT, and the EH hold simultaneously and the distribution is Gaussian~\cite{Lapas2007}; this dominates the diffusive band $0<\alpha<2$. Partial relaxation, $0<R_{\text{ext},\infty}\leq 1$, signals broken ergodicity and long-lived correlations, with a narrow transition near $\alpha \simeq 2$ with $\beta > 0$. For $R_{\text{ext},\infty}>1$, the drive-induced drift overruns the
initial scale $\langle A^2(0)\rangle$, marking the runaway sector (specially green region in Fig.~\ref{fig:beta_alfa_z}(c)).

The asymptotics of Eq.~\eqref{eq:limR(t)} separate the drive (numerator) from internal dissipation (denominator). For $\beta>0$ the drive vanishes as $z^{\beta}\to0$, leaving an $\alpha$-only result: $R_{\text{ext},\infty} = 1$ for $\alpha>2$ (hyperballistic, yellow in Fig.~\ref{fig:beta_alfa_z}(c)), $R_{\text{ext},\infty} = (1+c_T)^{-1}$ for $\alpha=2$ (ballistic), and $R_{\text{ext},\infty} = 0$ for $\alpha<2$ (diffusive). For $\beta=0$ the force merely shifts these values to $1+\epsilon k$, $(1+\epsilon k)/(1+c_T)$, and $0$, respectively. The decisive case is $\beta<0$, where $z^{\beta}$ diverges and $z\tilde{R}_{\text{ext}}(z)\sim (\epsilon k/c_T)z^{\beta-\alpha+2}$. This defines the universal control parameter for diffusive regimes:
\begin{equation}
    \eta = \beta - \alpha + 2,
    \label{eq:eta}
\end{equation}
where $\eta<0$ makes $R_{\text{ext}}$ diverge (non-thermal steady states with persistent drift), $\eta=0$ balances dissipation and forcing, and $\eta>0$ fully dissipates the drive and restores equilibrium. The single scaling law $\eta$ thus unifies the diffusive, hydrodynamic, and force-dominated regimes.

{\it Evolution of the temperature from the relaxation function. ---}\label{sec:evol-temp-func-resp}
The ensemble averages of $A$ and $A^2$ follow from Eqs.~\eqref{eq:GLE} and~\eqref{eq:Rext}:
\begin{equation}
    \langle A(t) \rangle = \langle A(0) \rangle \left[ \kappa\,\Delta R(t) + R_{\text{ext}}(t) \right],
    \label{eq:A(t)_average}
\end{equation}
\begin{align}
    \langle A^2(t) \rangle &= D_T \big[1 - R^2(t)\big] \notag \\
    &\quad +\langle A^2(0) \rangle \big[\kappa\,\Delta R^2(t) + R_{\text{ext}}^2(t)\big],
    \label{eq:<A2(t)>_tot}
\end{align}
with $\kappa = \langle A^2(0) \rangle / \langle A(0) \rangle^2 - 1$ and $\Delta R(t)=R_{\text{ext}}(t)-R(t)$. The FVT gives $\lim_{t\to\infty}\langle A(t)\rangle = \lim_{z\to 0} z[\langle A(0)\rangle + \tilde{F}_{\text{ext}}(z)]\tilde{R}(z)$, which may retain a residual drift. Identifying $\lim_{t\to\infty}\langle A^2(t)\rangle = m k_B T_{\text{ef,st}}$ and using Eqs.~\eqref{eq:morgado} and~\eqref{eq:asympF},
\begin{align}
T_{\text{ef,st}} &= T_R + \lim_{z \to 0} \frac{1}{\left(1+c_T z^{\alpha-2}\right)^2} \times  \nonumber \\
&\quad \left\{T_0 - T_R + \frac{k z^{\beta}}{m k_B} \left[2\langle A(0)\rangle + k z^{\beta}\right]\right\}.
\label{eq:<A2(z)>_tot_lim2}
\end{align}
The constant $D_T$ is fixed by the noise correlator, Eq.~\eqref{eq:C_xi-1};
the FDT in the absence of forcing identifies it as $D_T = m k_B T_R$,
and the initial second moment defines $T_0 \equiv \langle A^2(0)\rangle/(m k_B)$. The interplay of $\alpha$ and $\beta$ then determines whether
$T_{\text{ef,st}}$ relaxes to $T_R$ or settles at a different value: the
drive-dependent term $k z^{\beta}$ acts as a non-equilibrium contribution
that competes with the relaxation set by $\tilde{\Pi}(z) \sim c_T z^{\alpha-1}$,
and when it fails to vanish as $z \to 0$, the system sustains a steady
state with $T_{\text{ef,st}}\neq T_R$ supported by a balance between the
work injected by the drive and the energy dissipated into the bath.

{\it Generalized Fluctuation--Dissipation Theorem. ---}\label{sec:gFDT}
From Eqs.~\eqref{eq:A(t)_average} and~\eqref{eq:<A2(t)>_tot}, the variance
of $A(t)$ reads
\begin{equation}
\sigma_A^2(t) = \sigma_A^2(0)\,R^2(t)+ D_T \left[1-R^2(t)\right],
\label{eq:sigma_A2}
\end{equation}
so that $\sigma_A^2(t) = \sigma_{A^\dagger}^2(t)$, i.e. the fluctuations of
the driven and undriven variables coincide, confirming that the drive
enters only through the mean response. In the stationary limit,
Eq.~\eqref{eq:sigma_A2} yields
\begin{equation}
D_T = \frac{\sigma^2_{A,\text{st}}-\sigma_A^2(0)\,R_\infty^2}{1 - R_\infty^2},
\label{eq:C_M}
\end{equation}
a consistency relation between $D_T$, the stationary variance, and the
asymptotic relaxation. When KT holds
($R_\infty = 0$, $0<\alpha<2$), Eq.~\eqref{eq:C_M} collapses to the
standard FDT, $C_\xi(t) = 2 m k_B T_R\,\Pi(t)$. Otherwise, the relation
generalizes to $C_\xi(t-t') = 2 m k_B D_T\,\Pi(t-t')$ with $D_T$ now
constrained by $R_\infty$, extending Kubo's result to the full
non-Markovian regime. Substituting Eq.~\eqref{eq:limR-2} gives
\begin{equation}
D_T = \frac{\sigma^2_{A,\text{st}}-\sigma_A^2(0)\,(1+M_{\omega})^{-2}}
            {1 - (1+M_{\omega})^{-2}},
\label{eq:CM-ass}
\end{equation}
which admits three limits. For $M_\omega \to \infty$, $R_\infty \to 0$
and correlations with the initial state are erased, thus the standard FDT is
recovered. For finite $M_\omega$, partial memory survives and the system
sustains intermediate effective temperatures consistent with
Eq.~\eqref{eq:<A2(z)>_tot_lim2}. For $M_\omega \to 0$ (the hyperballistic
limit), $R_\infty \to 1$ and Eq.~\eqref{eq:CM-ass} becomes singular:
no constant $D_T$ can be defined, KT fails, and the FDT
admits no formulation (signatures of genuinely non-ergodic dynamics).


{\it Numerical verification. ---}\label{sec:num_results} The $\eta$-classification admits a stringent, parameter-free test in the analytically tractable Markovian limit, i.e. the normal-diffusion fixed point of Eq.~\eqref{eq:morgado} ($\alpha=1$, $c_T=\gamma$), for which $R(t)=e^{-\gamma t}$, $R_\infty=0$, and KT holds. The control parameter reduces to
$\eta=\beta+1$, and for $\langle A(0)\rangle=0$ Eq.~\eqref{eq:<A2(z)>_tot_lim2} collapses to
\begin{equation}
T_{\text{ef,st}} = T_R + \frac{k^{2}}{m k_B\,\gamma^{2}}\,
\lim_{z\to0} z^{\,2\eta}
=\begin{cases}
T_R, & \eta>0,\\[3pt]
T_R+\dfrac{k^{2}}{m k_B\,\gamma^{2}}, & \eta=0,\\[5pt]
\infty, & \eta<0,
\end{cases}
\label{eq:Tef_markov}
\end{equation}
assigning thermalization, a finite non-thermal plateau, and runaway heating to the three sectors. Equation~\eqref{eq:Tef_markov} also exposes the selective ergodicity breaking explicitly: defining the drive-induced drift $\Phi(t) \equiv \int_0^t e^{-\gamma(t-t')}F_{\text{ext}}(t')\,dt' = \langle p(t)\rangle$, the exact second moment factorizes as $\langle p^2(t)\rangle = \Phi^2(t) + \sigma_p^2(t)$, with $\sigma_p^2(t) = m k_B T_R + m k_B(T_0 - T_R)\,e^{-2\gamma t}$ strictly drive-independent, in accord with Eq.~\eqref{eq:sigma_A2}. The drive feeds the mean alone; the variance never knows it is there.

\begin{figure}[t]
\centering
\includegraphics[width=0.95\linewidth]{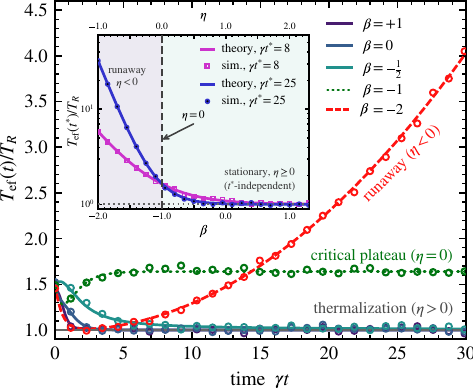}
\caption{Normalized effective temperature in the Markovian limit
($\alpha=1$, $\eta=\beta+1$). \textbf{Main panel:}
$T_{\text{ef}}(t)/T_R = \langle p^2(t)\rangle/(m k_B T_R)$ for five
drives sharing the low-$z$ scaling $\tilde{F}_{\text{ext}}(z)\sim k z^{\beta}$
but differing in functional forms: $F_{\text{ext}}(t) = 1.2(e^{-t}-3e^{-3t})$
($\beta\!=\!+1$), $1.35\,e^{-1.5t}$ ($\beta\!=\!0$), $0.6\,t^{-1/2}$
($\beta\!=\!-1/2$), $F_0\!=\!0.8$ (constant, $\beta\!=\!-1$), and
$0.06\,t$ ($\beta\!=\!-2$); reduced units $\gamma=m=k_B=T_R=1$.
Open symbols: Langevin simulations ($\sim\!10^4$ trajectories); lines:
exact analytic result. \textbf{Inset:} continuous sweep in $\beta$
(top axis: $\eta$) for $F_{\text{ext}}(t)=F_0(1+t/\tau)^{-(\beta+1)}$
with $F_0\!=\!0.8$, $\tau\!=\!4$, recorded at $\gamma t^{*}=8$ and $25$.
Curves collapse for $\eta\geq0$ (steady state) and separate for $\eta<0$
(runaway), crossing precisely at $\eta=0$. Common parameters: $T_R=1$,
$T_0=1.5$, $\gamma=1$.}
\label{fig:Tsim}
\end{figure}

We integrate $dp(t)/dt = -\gamma p(t) + \xi(t) + F_{\text{ext}}(t)$ with
$\langle \xi(t)\xi(t') \rangle = 2\gamma m k_B T_R\,\delta(t-t')$ using
the exact Ornstein--Uhlenbeck propagator (force held constant over each
step). This scheme is free of discretization bias and recovers
$\langle p^2\rangle = m k_B T_R$ exactly at equilibrium, a prerequisite
for resolving the small departures predicted by
Eq.~\eqref{eq:Tef_markov}. Crucially, for $\eta>0$
Eq.~\eqref{eq:Tef_markov} pins $T_{\text{ef,st}} = T_R$ for
\emph{any} amplitude and functional force form, since the prefactor $k$ multiplies $z^{2\eta}\to 0$; the chosen constants merely separate the curves.
The sole amplitude-sensitive prediction is the plateau $T_R + (F_0/\gamma)^2$ at $\eta=0$.

Figure~\ref{fig:Tsim} confirms the trichotomy of Eq.~\eqref{eq:Tef_markov} within statistical error: The three $\eta>0$ forces relax to $T_R$ independently of its detailed temporal profile, the constant drive saturates at the predicted plateau $T_R+(F_0/\gamma)^2=1.64$, and the linear ramp grows without bound. The inset isolates the transition itself, i.e. sweeping $\beta$, the two observation times $\gamma t^{*}=8$ and $25$ collapse onto a single time-independent profile for $\eta\geq 0$ (the operational signature of a true steady state) and separate for $\eta<0$, the hallmark of runaway. Their crossing precisely at $\eta=0$ furnishes a parameter-free confirmation of the scaling criterion, anchoring the full $(\alpha,\beta)$ diagram of Fig.~\ref{fig:beta_alfa_z}.

{\it Conclusions. ---} The non-Markovian GLE under any external force whose Laplace transform
exhibits a power-law tail at small $z$ reveals a universal phase diagram
governed entirely by the single parameter $\eta=\beta-\alpha+2$, which
balances internal memory ($\alpha$) against the spectral content of the
forcing ($\beta$): thermalization ($\eta>0$), non-ergodic saturation at a
finite plateau ($\eta=0$), and force-dominated runaway ($\eta<0$). This
framework encompasses the regimes of anomalous diffusion, including the
ballistic case, and provides unified control over diverse transport
phenomena in non-equilibrium systems. The FDT survives in an extended
form valid throughout the range $0<\alpha\leq 2$, with the effective
coupling $D_T$ fixed by the spectral integral $M_\omega$; ergodicity
breaks selectively, with $\sigma_A^2(t)$ strictly drive-independent while
$\langle A(t)\rangle$ encodes the drive in full. Markovian Langevin
simulations confirm the $\eta$-classification quantitatively, reproducing
both the critical plateau and the time-independence that distinguishes a
steady state from runaway. Since the criterion involves only asymptotic
exponents, it is experimentally actionable: tuning the long-time tail of
a drive suffices to steer a memory-laden system across the
ergodic--nonergodic boundary, with direct relevance to viscoelastic
media, active matter, and driven open quantum systems. By tuning
ergodicity breaking through a single external parameter, our results
provide a versatile route to controlling transport in non-equilibrium
systems.

{\it Acknowledgments. ---} L.C.L. gratefully acknowledges financial support from the National Institute of Science and Technology in Innovative Research in Health Sciences – from Nanotechnology to Artificial Intelligence (INCT PICS), sponsored by the Brazilian National Council for Scientific and Technological Development (CNPq), grant no. 408417/2024-2. R.V.B.M. acknowledges support from the Coordination for the Improvement of Higher Education Personnel (CAPES). M.F.C. thanks PIP CONICET 11220200101599CO and UNGS-80020250100042GS.

\bibliography{references}

\end{document}